\documentclass[aps,pre,twocolumn,showpacs,floatfix,preprintnumbers,superscriptaddress,amsmath,amssymb]{revtex4-1}
\usepackage{amsmath}
\usepackage{amssymb}
\usepackage{graphicx}
\usepackage{epsfig}
\usepackage{dcolumn}
\usepackage{bm}
\usepackage{array}
\usepackage{mathtools}
\usepackage[colorlinks,bookmarks=false,citecolor=blue,linkcolor=red,urlcolor=blue]{hyperref}

\begin{document}
\title{Diffraction of a CW atom laser in the Raman-Nath regime}
\author{Sumit Sarkar}
\author{Jay Mangaonkar}
\author{Chetan Vishwakarma}
\affiliation{Department of Physics, Indian Institute of Science Education and Research, 
Dr. Homi Bhabha Road, Pune 411 008, India}
\author{Umakant D. Rapol}
\affiliation{Department of Physics, Indian Institute of Science Education and Research, 
Dr. Homi Bhabha Road, Pune 411 008, India}
\affiliation{Center for Energy Sciences, Indian Institute of Science Education and Research, 
Dr. Homi Bhabha Road, Pune 411 008, India}

\date{\today}
\begin{abstract}

Atom interferometry is the most successful technique for precision metrology. However, current interferometers using ultracold atoms allows one to probe the interference pattern only momentarily and has finite duty cycle, resulting in an aliasing effect and a low-bandwidth measurement -- also known as Dick effect. Interferometry with a continuous-wave atom  laser shows promise in overcoming these limitations due a continuous monitoring of the interference pattern. In this work, we demonstrate a key step towards such an interferometry by demonstrating a diffraction of an `atom laser' in the Raman-Nath regime. We outcouple a continuous beam of coherent atoms from a reservoir of $^{87}$Rb Bose-Einstein condensate (BEC) upto 400 ms. The `atom laser' interacts with a grating formed by a standing wave of a far detuned laser light. The atom laser diffracts into several orders going up to 9$^{th}$ order or up to momenta of $\pm 18\ \hbar k$. We have characterized the diffraction of atom laser for different conditions and the results match with numerical simulations. Such atom laser will allow for construction of an atom-interferometer to probe physics phenomenon continuously up to a time of the order of few hundred millisecond.

\end{abstract}

\maketitle

Atom interferometry is a matter-wave based interferometer realized through coherent manipulation of translational and internal degrees of freedom of atoms or molecules \cite{alex}. Atoms/molecules possess several physical properties (e.g. magnetic moment, mass, high collision cross-section, polarizibility etc.) which enables them to interact with various environments (e.g. Magnetic fields, gravity, electric fields etc.) making the interferometer highly sensitive to the quantity of interest as opposed to photon based interferometers. Atom-interferometry has gained impetus in studying fundamental quantum science, precision metrology \cite{Bongsk, matt} and for next generation quantum technologies \cite{bongs}. Atom interferometry has been used to measure rotations \cite{lenef, gaustav}, fine structure constant $\alpha$ \cite{gupta, bouch, wicht}, local gravity \cite{kase, peter, geiger}, gravity gradients \cite{snad, guirk}, atomic polarizability \cite{eks} etc. with unprecedented accuracy and precision. In addition, precision metrology based on atom-interferometry has provided world-wide frequency standards\cite{nichol, ian, derev}. The most popular atom-interferometers are based on thermal or cold atomic samples. A thermal beam of atoms provide sensitivity to measure a phase shift of $\sim 10^{-3}$ rad \cite{alex}. Whereas, an ultracold cloud of BEC can further increase the sensitivity by a factor of $10^3$ \cite{alex}. In spite of the advantages mentioned above, these interferometers cannot be used to monitor a physical process continuously, as in most of the cases the giant matter waves involved in the process interfere for a very short time and provide a one-shot measurement of the phase shifts. For monitoring a slowly varying physical parameter, the measurement needs to be repeated many times. This measurement process gets severely limited by the duty-cycle of production of the atomic wave packets. Therefore, such measurements are limited in bandwidth and suffer from aliasing effect \cite{fang}, also known as Dick effect \cite{dick}. A possible solution is to have a form of matter wave which is: ({\it a}) Continuous source of the interfering wavepackets (similar to a laser beam) and ({\it b}) maintaining long coherence times. These properties can be achieved with a continuously outcoupled stream of ultracold atoms from a reservoir. One such source is the outcoupling of coherent beam of ultracold atoms from a reservoir formed of a BEC. This outcoupled beam of atoms, possessing a long coherence length, high collimation and high brightness (limited by the flux) is known as ‘atom laser’ \cite{mewes}.
Because, of better spatial-mode properties and less wavefront aberrations \cite{robins} atom lasers are the most desired candidate for precision measurements of inertial effects. 
\\
There has been significant amount of work over a couple of decades regarding generation and characterization of atom lasers. Several techniques have been used to generate atom lasers in a pulsed or continuous form, such as : ({\it i}) radio frequency outcoupling \cite{mewes, bloch, martin, coq}, ({\it ii}) outcoupling using Raman beams \cite{hagley, rob}, ({\it iii}) outcoupling using spilling from a optical or hybrid trap (optical dipole trap $+$ quadrupolar magnetic trap) \cite{cenn}, and ({\it iv}) using guided push beam\cite{guerin, bernard}. Realization of a pumped atom laser has been demonstrated in Ref. \cite{nicholas}. Using an atom laser, beam splitter and Bragg mirrors have been demostrated \cite{gatto, fabre}. Measurement of coherence properties like temporal coherence and second order correlation function ($g^{(2)}$) have also been measured for an atom laser \cite{mich, anton}.
 In Ref. \cite{zoran}, interference due to an overlap of many matter waves leaking from many sites of an optical lattice has been demonstrated. In Ref. \cite{anderson} the authors demonstrated a pulsed atom laser formed due to interference of many Airy functions from a tilted optical lattice.  But, to the best of our knowledge, no attempt has been made so far in setting up an interferometer with atom laser. The first step towards atom-interferometry based on atom laser would be to do a coherent splitting of the atom laser and interfere the different atom lasers from different parts. We demonstrate diffraction using an atom laser which would open up a new direction towards building such an interferometer. In this work, we have demonstrated the diffraction of an atom laser in the Raman-Nath regime \cite{raman}. This configuration can be easily modified to set up a standard three-grating interferometer \cite{rasel} using atom laser.
\\
In our work, we have successfully demonstrated Kapitza-Dirac effect \cite{kapitza} using a continuous atom laser extracted from a reservoir of BEC. Amongst all the known techniques to outcouple an atom laser from a reservoir in CW or pulsed form, we choose the spilling method to generate a CW atom laser for several reasons : ({\it i}) a well-collimated atom laser can be achieved even without guiding, ({\it ii}) the flux rate can be controlled precisely and ({\it iii}) the acceleration in the direction of gravity can be reduced by applying a magnetic field gradient. To study the diffraction of atom laser a pulsed light-grating is placed in the path of falling atom laser. We use square pulse switching of the light-grating with different pulse-width to study the interaction of atom laser with the light-grating. Here we briefly summarize the relevant theory to understand the phenomenon. The light-grating is made of a 1-D optical lattice with electric field given by:
\begin{equation}
E(x,t) = 2E_0\cos{(kx)}\cos{(\omega t)}
\end{equation}
\begin{figure}[t]
\includegraphics*[width=3.0in]{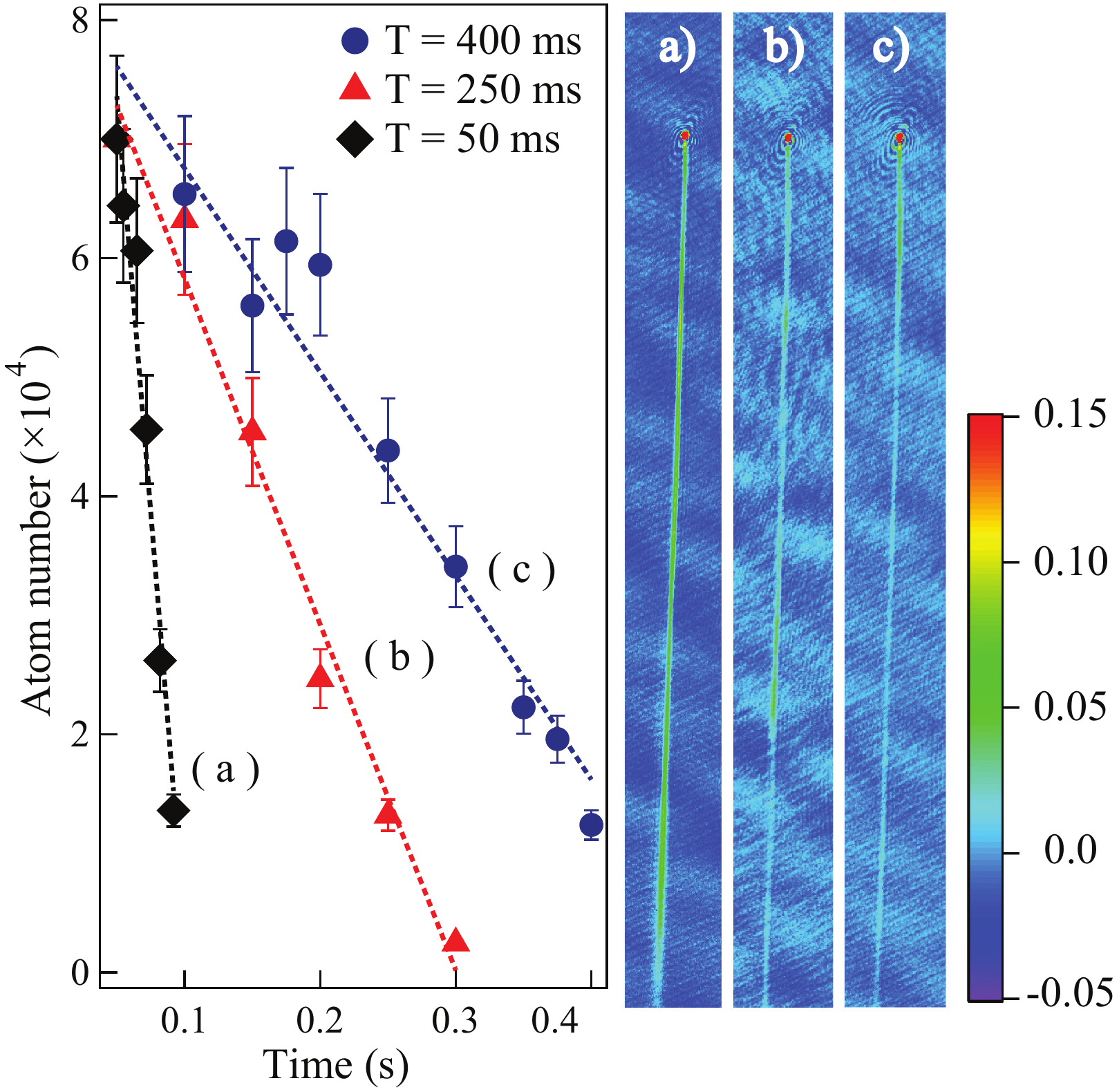}
\caption{(Color online) {\bf Left panel} : No of atoms in the reservoir as a function of time for three different outcoupling rates. Atomic flux for the three cases are: (a) $1.39\times10^6$ atoms/s, (b) $2.91\times10^5$ atoms/s, (c) $1.71\times 10^5$ atoms/s. {\bf Right panel} : Images of the atom laser corresponding to the three flux rates in the left panel. These images are taken at (a) $40$ ms, (b) $200$ ms and (c) $300$ ms after the start of outcoupling. Color scale represents optical density.}
\label{flux}
\end{figure}
\begin{figure}[b]
\includegraphics*[width=2.8 in]{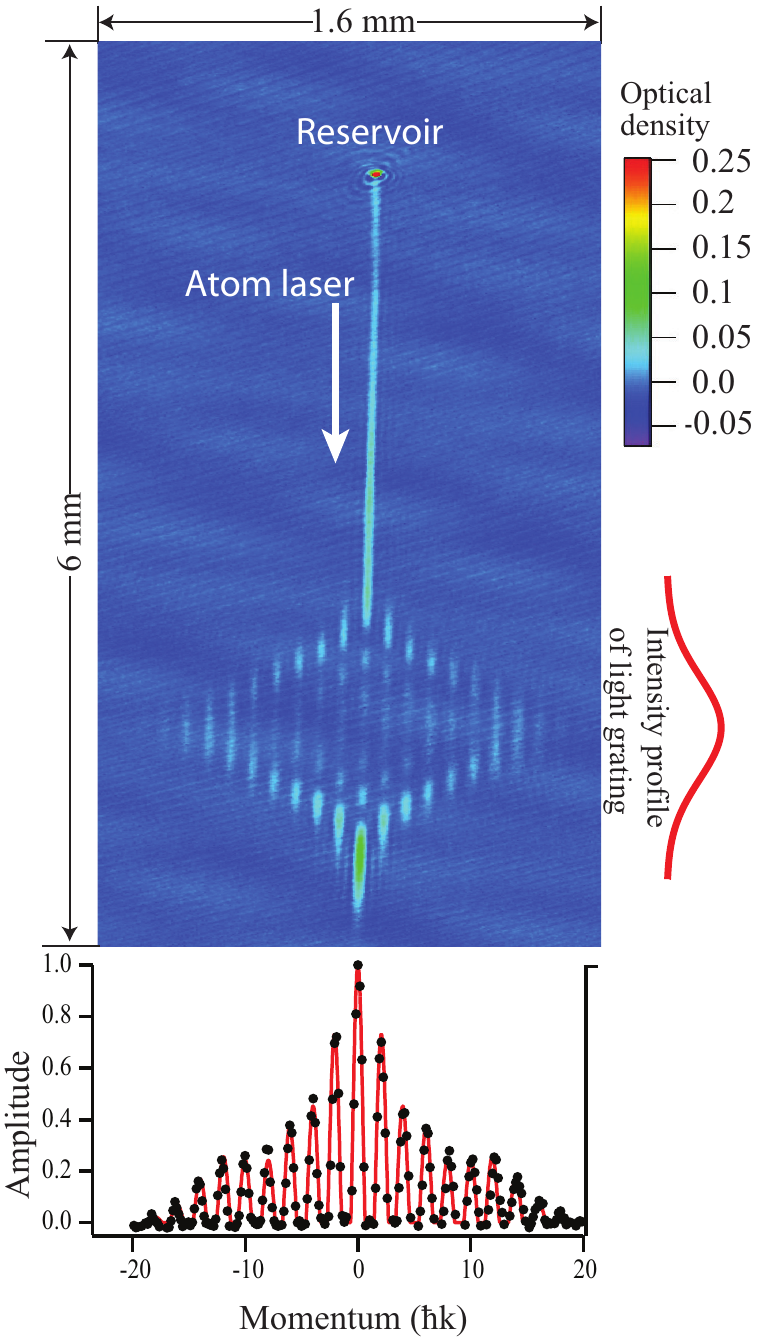}
\caption{(Color online)
The image in the upper panel shows a typical time of flight absorption image of the atom laser diffraction. The red curve drawn beside the image is the Gaussian intensity profile of the standing wave of light that forms the diffraction grating with which the atom laser interacts. The lower panel shows the normalized population in different diffraction orders (upto $\pm 18 \ \hbar$k). Circles represent experimental data whereas the solid line shows the numerical simulation. }
\label{schematic}
\end{figure}
The strength of the interaction can be defined using traveling wave Rabi frequency $\Omega = \mu E/\hbar$, where $\mu$ is the dipole matrix element between the two atomic states coupled by light. Assuming that the atom laser has negligible momentum in the direction of the lattice, we can understand the effect of grating as diffraction of the atoms into several momentum eigenstates. The momentum of each eigenstate can be labeled as a multiple of $2\hbar k$, such that, atoms in the $n^{th}$ diffracted order will possess $2n\hbar k$ momenta. Here $k=2\pi /\lambda$ represents the wavenumber of the lattice laser. The population $P_n$ in the $n^{th}$ diffraction order can be expressed as 
\begin{equation}
P_n = J_n^2(\theta); \hspace*{5mm} n = 0,\pm 1, \pm 2,...
\label{population}
\end{equation}
where, $J_n$ is the $n^{th}$ order Bessel function of the 1$^{st}$ kind and $\theta$ is the pulse area for the corresponding interaction time $\tau$. $\theta$ can be expressed in terms of peak Rabi frequency $\Omega_0$ and $\tau$ as:
\begin{equation}
\theta = \frac{\Omega_0^2 \tau }{2 \delta}
\end{equation}
where, $\delta = (\omega_l - \omega_a)$ is the detuning of the laser with $\omega_l$ being the laser frequency and $\omega_a$  the atomic resonance frequency. 
The condition for Kapitza-Dirac diffraction comes from the Raman-Nath approximation. The basic assumption underlying the theory is that the interaction time should be such that the distance travelled by the atoms during the interaction time is very less compared to the wavelength of lattice. Thus, the limiting pulse width is given by $\tau \ll 1/\omega_{rec}$, where $\omega_{rec} = \hbar k^2/2m$ is the recoil frequency. Following the argument given in ref. \cite{subha}, a more precise condition for Kapitza-Dirac scattering is given by $\tau < \tau_{osc}/4$, where - 
\begin{equation}
\tau_{osc} = \frac{\pi}{\Omega_0}\left( \frac{\mid\delta\mid}{\omega_{rec}}\right)^{1/2}
\label{condition}
\end{equation}
$\tau_{osc}$ is the period of oscillation of the atoms due to the curvature of the optical potential at the center of each lattice site.
\\
\begin{figure*}[t]
\begin{center}
\includegraphics[width=7in]{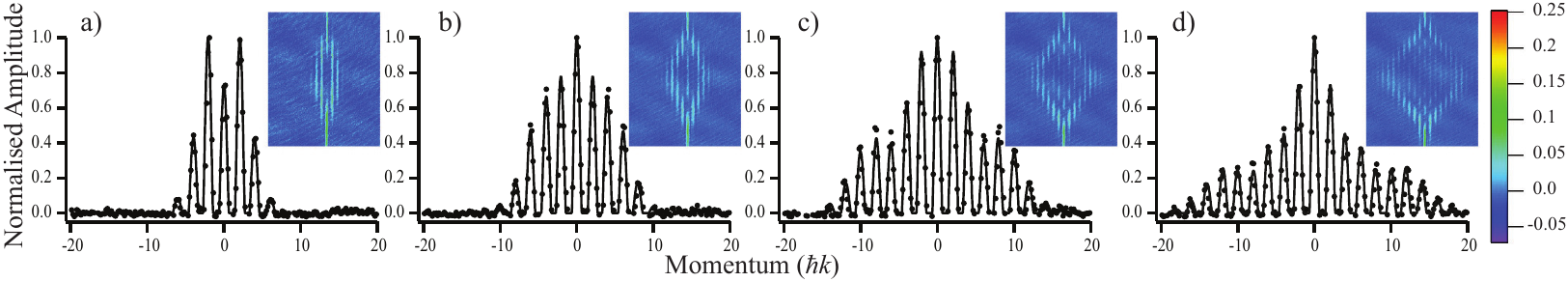}
\end{center}
\caption{(Color online) Normalized population in different diffraction orders with varying interaction time. (a) $\tau$ = 0.15\ $\mu s$, (b) $\tau$ = 0.3$\ \mu s$, (c) $\tau$ = 0.4\ $\mu s$ and (d) $\tau$ = 0.5\ $\mu s$, keeping the peak Rabi frequency and detuning constant. Black circles represents experimental data. Solid line represent numerical simulation for (a) $\theta$ = 2.64, (b) $\theta$ = 4.29, (c) $\theta$ = 6.62, (d) $\theta$ = 8.23. Insets show corresponding time of flight absorption images of the diffracted atom laser. The dimension of each image frame is $\sim$ 1.85 mm $\times$ 1.54 mm. Color scale represents the optical density.}
\label{diffraction}
\end{figure*}
The basic experimental set-up to realize the system is similar to that given in Ref \cite{sumit}. We load a standard magneto-optical trap (MOT) with $\sim$ 10$^7$ atoms of $^{87}$Rb. With laser cooling we reach upto $\sim$30 $\mu$K \cite{sunil} temperature. The atoms are then transferred into a crossed optical dipole trap (wavelength $\lambda=1064$ nm). Further evaporative cooling is performed by reducing the intensity of the lasers of the dipole trap to reach the quantum degeneracy level. We produce BEC with $\sim 7\times 10^4$ atoms. Starting from the loading of  MOT, the quadrupolar magnetic field is kept at constant axial gradient of $\sim $ 24.5 Gauss/cm. \\
To outcouple the atoms from the reservoir of BEC, the trap-depth of the crossed dipole trap is further reduced adiabatically keeping the magnetic field gradient constant. The flux of the atom laser is controlled by the rate of  lowering of the depth of optical dipole trap. Whereas, the axial field gradient helps us to reduce the net acceleration felt by the falling atoms. For the value of the field gradient used in our experiment, we could slow the atom laser down to an acceleration of $\sim$ g/5, where `g' is the acceleration due to gravity. In our experiment, keeping the magnetic field gradient constant, we have been able to outcouple atom laser with varying flux rates. Three different rates of outcoupling have been represented in Fig. \ref{flux}. The absorption images shown in Fig. \ref{flux} are taken at (a) 40 ms, (b) 200 ms and (c) 300 ms respectively after the start of outcoupling. To increase the signal-to-noise ratio (SNR), each of these images have been averaged over 15 realizations for the same set of experimental parameters. The three images correspond to three different outcoupling rates (a) 1.39 $\times$ 10$^6$ atoms/s, (b) $2.91\times 10^5$ atoms/s, (c) 1.71 $\times$ 10$^5$ atoms/s. The rate of outcoupling is measured by tracking the number of atoms left in the reservoir (see left panel of Fig. \ref{flux}). For the three cases of outcoupling shown in Fig. \ref{flux}, we could see the atom laser for a duration of (a) 50 ms, (b) 250 ms and (c) 400 ms respectively, before the reservoir is emptied.\\

To realize the Kapitza-Dirac effect with the above atom laser, a far detuned 1-D optical lattice is kept $\sim$ 3 mm below the reservoir. The lattice laser is $-6.8$ GHz detuned from 5$S_{1/2}F=1 \longrightarrow 5P_{3/2} F=2$ transition of $^{87}$Rb. The lattice is made of a collimated laser beam [waist($\omega_d$) $=$ 720 $ \mu$m]. Once the outcoupled atom laser falls enough to reach up to the position of the optical lattice, the lattice is turned on for a short time $\tau$. After the interaction, a free evolution time (time of flight) is given to resolve all the diffracted orders and an absorption image is taken. Fig. \ref{schematic} shows a typical time of flight image of the diffraction of an atom laser for $\tau =$ 0.5 $ \mu$s. A red Gaussian envelope has been shown in Fig. \ref{schematic} to schematically represent the overlapping region of the atom laser with the lattice beam. Due to the Gaussian profile of the optical lattice beam, different sections of the atom laser see a different intensity in the vertical direction. So the strength of the diffraction is maximum at the center of the lattice beam and falls as we move away from the center of the laser beam. The bottom panel in Fig. \ref{schematic} shows the normalized population in different momentum eigenstates due to the diffraction of an atom laser. To find out the population in the diffraction orders, we choose the section of the image where the effect of the grating can be observed. The size of the interaction region was found to be $\sim 2 \ \omega_d$, which includes $\sim 95 \ \%$ of the total power of the lattice beam. To find out the normalized population from the selected region we do a column integration of the optical density and divide it by the maximum population. In the lower panel of Fig. \ref{schematic}, the black circles represent the experimental data of normalized population for the diffraction pattern shown in the upper panel of the same figure. Whereas, the solid lines display the normalized population calculated numerically.\\

A broader picture of the interaction of the atom laser with the light-grating has been displayed in Fig. \ref{diffraction}. In this figure, we have represented diffraction of an atom laser by varying the interaction time. For our experimental parameters, the recoil frequency is found to be $\omega_{rec}=$ 23.68 KHz. Therefore, to satisfy the Raman-Nath condition for thin grating, the interaction time has to be less than 42 $\mu$s. Also, the condition for transition to classical oscillation regime can be found using Eqn. \ref{condition}. For the 1-D lattice used in our experiment, the peak value of travelling wave Rabi frequency is $\sim$ 470 MHz. Hence, the upper limit of the interaction time to observe Kapitza-Dirac diffraction is given by $\tau_{osc}/4 \sim $ 0.9 $ \mu$s. The values of the interaction times chosen in our experiment are (a) $\tau$ = 0.15 $\mu$s, (b) $\tau$ = 0.3 $\mu$s, (c) $\tau$ = 0.4 $\mu$s and (d) $\tau$ = 0.5 $\mu$s. Fig. \ref{diffraction} displays how the distribution of population in different momentum eigenstates evolves with increasing value of the interaction time. For the maximum value of interaction time ($\tau =$ 0.5 $\mu$s) chosen in this experiment, we have been able to transfer maximum momenta of $\pm$ 18$\hbar$k to the atoms. In Fig. \ref{diffraction}, the black circles represent the experimental data for normalized population in different diffraction orders. The solid lines represent numerically simulated population in diffracted orders for the values of $\theta$ approximately equal to that used in the experiment. The population in various diffraction orders are calculated numerically by convoluting the Thomas-Fermi distribution with Eqn. \ref{population}. The average Thomas-Fermi radius of the atom laser used to demonstrate diffraction is found to be $\sim$ 23 $\mu$m. The values of $\theta$ for which we found best fit to the diffracted population agrees with the values of $\theta$ used in the experiment within $\pm 10\%$ uncertainty. \\

In summary, we have demonstrated diffraction of a slowed down CW atom laser in Kapitza-Dirac limit. The atom laser can be outcoupled with well-controlled flux rate. In our experiment, the atom laser could be seen up to 400 ms for the lowest outcoupling rate used. Since for a constant outcoupling rate the time for which one can extract atom laser is proportional to the initial atom number in the reservoir, improving the number in the BEC will essentially allow us to use the atom laser for a longer time. We have numerically simulated the population distribution in the diffracted orders. Our experimental results are in good agreement with the numerical simulations. The extension of this work is to realize a three grating interferometer using atom laser which can be used for quasi-continuous probing of physical quantities with high bandwidth and improved sensitivity. In addition, one can use the schemes shown in Ref. \cite{horik, daniel} to produce BEC on an atom chip with a fast production rate and have multiple of those reservoirs on a chip that operate in a synchronised manner to produce a near perpetual source of atom lasers that can be integrated with this diffraction effect for producing an interferometer that operates continuously.  

\begin{acknowledgments}
The authors would like to thank the Department of Science and Technology, Govt. of India 
for funds through grant no. EMR/2014/000365 and the Council of Scientific and Industrial Research, Govt. of India for funds through grant no. 03(1378)/16/EMR-II. 
\end{acknowledgments}

\end{document}